\documentclass[aps,prl,amsfonts,floatfix,superscriptaddress,longbibliography,nofootinbib,twocolumn]{revtex4-2}
\usepackage[utf8]{inputenc}
\usepackage{amsmath}    
\usepackage{mathbbol}
\usepackage{graphicx}   
\usepackage{verbatim}   
\usepackage{color}      
\usepackage{subfigure}  
\usepackage{hyperref} 

\usepackage{siunitx}
\usepackage{physics}
\usepackage{soul}
\usepackage{enumerate}
\usepackage{empheq}
\usepackage{multirow}

\usepackage[dvipsnames,svgnames]{xcolor} 

\begin{document}

\title{Quantum sensing in Kerr parametric oscillators}

\author{Jorge Ch\'avez-Carlos}
\affiliation{Department of Physics, University of Connecticut, Storrs, Connecticut 06269, USA}

\author{Daniela Garrido-Ram\'irez}
\affiliation{Faculty of Sciences, University National Autonomus of Mexico,. Postal 70-543, C.P. 04510, Cd. Mx, Mexico}

\author{A. J. Vega Carmona}
\affiliation{Departamento de F\'isica, Universidad Aut\'onoma Metropolitana-Iztapalapa, Av. Ferrocarril San Rafael Atlixco 186, C.P. 09310 CDMX, Mexico.}

\author{Victor S. Batista}
\affiliation{Department of Chemistry, Yale University, P.O. Box 208107, New Haven, Connecticut 06520-8107, USA}

\author{Carlos A. Trallero-Herrero}
\affiliation{Department of Physics, University of Connecticut, Storrs, Connecticut 06269, USA}

\author{Francisco P\'erez-Bernal}
\affiliation{Departamento de Ciencias Integradas y Centro de Estudios Avanzados en F\'isica, Matem\'aticas y Computaci\'on, Universidad de Huelva, Huelva 21071, Spain}
\affiliation{Instituto Carlos I de Física Teórica y Computacional,
  Universidad de Granada, Granada 18071, Spain}

\author{M. A. Bastarrachea-Magnani}
\affiliation{Departamento de F\'isica, Universidad Aut\'onoma Metropolitana-Iztapalapa, Av. Ferrocarril San Rafael Atlixco 186, C.P. 09310 CDMX, Mexico.}

\author{Lea F. Santos}
\affiliation{Department of Physics, University of Connecticut, Storrs, Connecticut 06269, USA}

\begin{abstract}
Quantum metrology and quantum sensing aim to use quantum properties to enhance measurement precision beyond what could be classically achieved. Here, we demonstrate how the analysis of the phase space structure of the classical limit of Kerr parametric oscillators can be used for determining control parameters values that lead to the squeezing of the uncertainty in position and the amplification of the quantum Fisher information. We also explore how quantum sensing can benefit from excited-state quantum phase transitions, even in the absence of a conventional quantum phase transition. The system that we consider models exciton-polariton condensates and superconducting circuits, making our study relevant for potential experimental applications.
\end{abstract}

\maketitle

The precise measurement of physical quantities is crucial for fundamental science and technological applications that range from navigation to telecommunications, timekeeping, and medical diagnostics. Quantum metrology and quantum sensing seek to take advantage of quantum properties to achieve measurement precision that surpasses classical limits~\cite{Cappellaro2017, Braun2018}. Entanglement, for example, is a pure quantum property that is harnessed to enhance  measurement sensitivity~\cite{D'Ariano2001, Acin2001, Giovannetti2001, Giovannetti2004, Giovanetti2006}. Likewise, squeezed light has been used to boost the sensitivity of laser interferometry-based detectors~\cite{Caves1981, Abadie2011}. Another promising approach is quantum criticality, which extends the use of phase transitions in classical sensing to the quantum domain~\cite{Zanardi2008}. At a quantum phase transition (QPT), the ground state's diverging susceptibility can, in principle, be exploited to enhance parameter estimation~\cite{Zanardi2008, Tsang2013, Fernandez2017, Rams2018, Garbe2020, Gietka2021, Chu2021, Salado2021, Ilias2022, Garbe2022, Ding2022, Candia2023, Hotter2024}. Experimentally, having access to external control parameters that can drive the system to critical points can be used for optimizing the performance of a sensing platform.

Complementing those existing studies, we introduce an approach for identifying the parameters values that enhance quantum sensitivity, that is based on the analysis of the phase-space topology of the system's classical limit. We focus on a Kerr parametric oscillator that can be realized with superconducting circuits~\cite{XiaoXu2023,Frattini2024} and an exciton-polariton condensate~\cite{Ghosh2020}. By changing the parameters of the system's Hamiltonian, we identify values that trigger the onset of a saddle point. To  accommodate the appearance of this point, the surrounding energy contours of the classical phase space get deformed. Simultaneously, the Wigner functions of the eigenstates associated with those energies get squeezed in position.

The system under investigation does not undergo a QPT, that is, the gap between the ground-state energy and the first-excited energy does not vanish. However, it does exhibit energy crossings at higher energy levels due to a phenomenon known as excited state quantum phase transition (ESQPT)~\cite{Caprio2008,Cejnar2021}. Despite growing interest in ESQPTs for their impact on localization~\cite{SantosBernal2015,Santos2016} and their applications to Kerr-cat qubits~\cite{Frattini2024, Chavez2023, Iachello2023}, this topic has received little attention in the fields of quantum metrology and quantum sensing. In systems that present both a QPT and an ESQPT, such as the Lipkin-Meshkov-Glick model~\cite{Chu2021, Garbe2022b} and Kerr parametric oscillators~\cite{Heugel2019,Candia2023,Asjad2023}, only their QPT has been explored for quantum sensing, leaving the potential of ESQPTs largely unexplored.

We show that the level crossings at excited states occur at specific values of the control parameters. These values are linked with changes in the classical phase-space structure that includes the emergence of the saddle point mentioned above. The crossings give rise to peaks in the quantum Fisher information (QFI) \cite{Braunstein1994,Braunstein1996}, a metric frequently employed to identify sudden changes in the nature of a quantum state. Interestingly, we demonstrate that the elevated values of the quantum Fisher information (QFI) are not limited to the excited states, but propagate down to the ground state. Additionally, we observe a direct correlation between the increased QFI values and the squeezing in position of the corresponding quantum states.

The parameters of Kerr parametric oscillators can be tuned in superconducting circuits~\cite{XiaoXu2023,Frattini2024} and, potentially, in exciton-polariton condensates~\cite{Ghosh2020}. This implies that our results could be experimentally detected and explored for sensing devices.

{\em Kerr parametric oscillator.---} The effective Hamiltonian of the Kerr parametric oscillator that we analyze is given by
\begin{equation}
\hat{H}=-\Delta\hat{a}^\dagger\hat{a}+K\hat{a}^{\dagger2}\hat{a}^2-P_0(\hat{a}+\hat{a}^\dagger),
\label{eq:Ham}
\end{equation}
where $\hbar=1$, $\hat{a}^\dagger$ ($\hat{a}$) is the creation (annihilation) operator, $\Delta$ is the detuning amplitude, $K$ is the Kerr nonlinearity, and $P_{0}$ is the squeezing amplitude. The Hamiltonian does not exhibit a QPT and the ground state does not present any singularity.

In superconducting circuits~\cite{Frattini2024,Kwon2022,He2023,Venkatraman2023}, the effective Hamiltonian in Eq.~(\ref{eq:Ham}) can be obtained by driving a SNAIL (superconducting nonlinear asymmetric inductive element) transmon~\cite{Frattini2017} with a frequency very close to the natural frequency of the oscillator~\cite{XiaoXu2023,Venkatraman2022}. In these platforms, the energy levels can be experimentally measured as a function of the control parameter~\cite{Frattini2024}. The three parameters, $\Delta$, $K$, and $P_0$, can be tuned through the drive and the hardware design. In the case of the exciton-polariton system proposed in~\cite{Ghosh2020}, the Hamiltonian in Eq.~(\ref{eq:Ham}) describes the low-lying spectrum of a condensate at zero momentum. It is formed inside a semiconductor microcavity and pumped with an external coherent optical field~\cite{Ciuti2005,Amo2016,Byrnes2014}, and the Kerr-like nonlinearities are potentially tunable~\cite{Carusotto2013}.

\begin{figure}[t]
\centering\includegraphics[width=1.0\columnwidth]{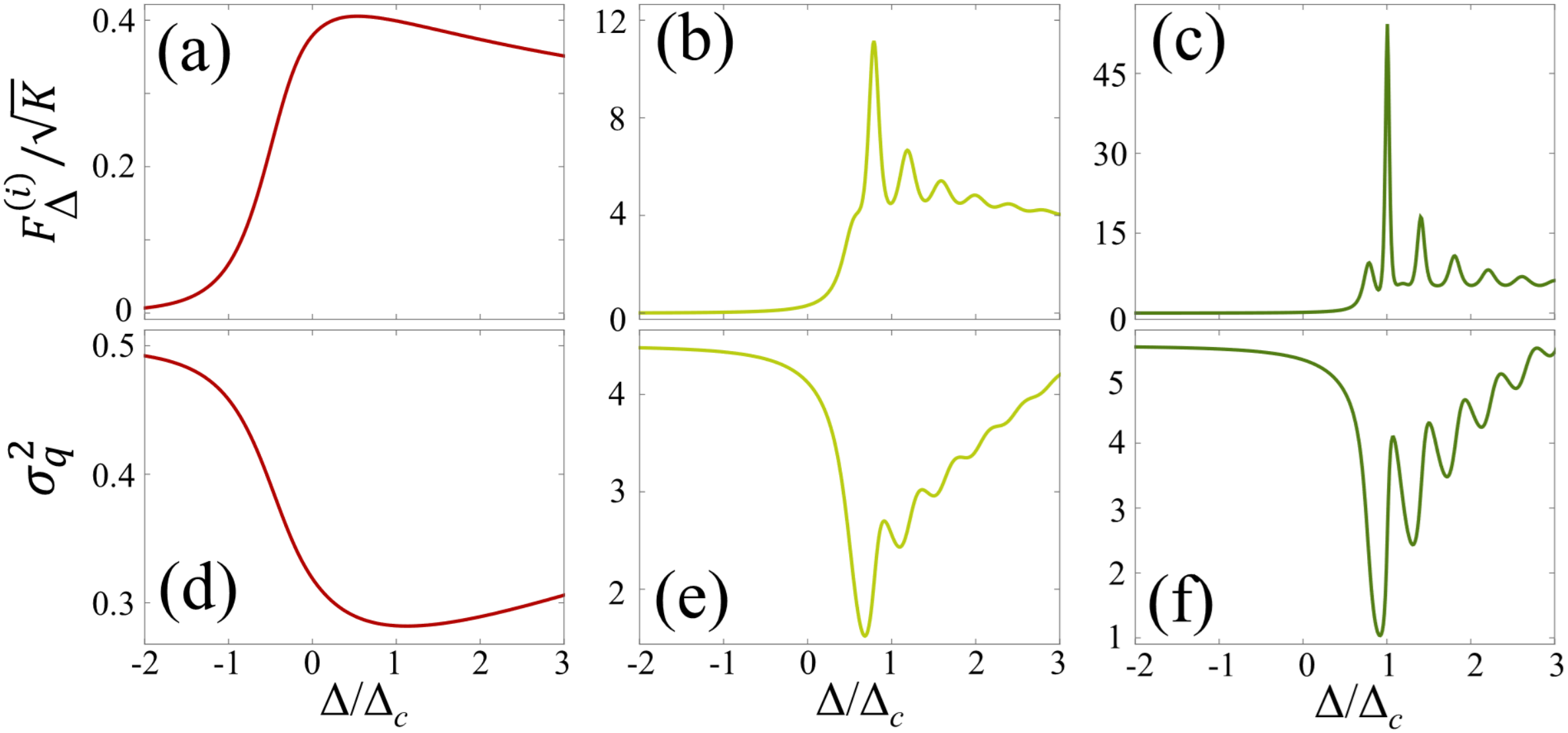}
	\caption{(a)-(c) Quantum Fisher information and (d)-(e) variance of position as a function of the control parameter $\Delta$ for the (a),(d) ground state, (b),(e) eigenstate $|\Psi_4\rangle $ and (c),(f) eigenstate $|\Psi_5\rangle $.}
		\label{fig:1}
	\end{figure}

{\em Quantum sensitivity enhancement.---} In Figs.~\ref{fig:1}(a)-(c), we set $P_{0}/K$ constant and study the QFI, defined as~\cite{Braunstein1994,Braunstein1996}
\begin{equation}
F^{(i)}_\Delta = {4 } \sum_{j\neq i}\frac{|\langle \Psi_{j}|\frac{\partial \hat{H}}{\partial \Delta}|\Psi_i\rangle|^{2}}{(E_{j}-E_{i})^{2}},
\label{eq:fisher}
\end{equation}
where $\hat{H}|\Psi_i \rangle = E_i |\Psi_i \rangle$ and $\frac{  \partial \hat{H} }{\partial \Delta }=-\hat{a}^\dagger\hat{a} = - \hat{n}$. The QFI is equivalent to the quantum fidelity susceptibility~\cite{You2007, Gu2010} and coincides with the real part of the quantum geometric tensor~\cite{Campos2007, Gu2010, Gietka2021}. The importance of this metric for quantum sensing becomes evident with the quantum Cram\'er-Rao bound, which states that the inverse of the QFI limits the accuracy of an unbiased estimation of a system parameter~\cite{Braunstein1994,Braunstein1996}.

The QFI for the ground state, $i=0$, diverges at a QPT, since $E_1-E_0 \rightarrow 0$. Our system does not exhibit a QPT, yet, we observe in Fig.~\ref{fig:1}(c) that $F^{(5)}_\Delta$ for the eigenstate $|\Psi_5\rangle $ shows a sharp peak when $\Delta \sim \Delta_c$. The value of $\Delta_c$ depends on $K$ and $P_0$ and will be explained below. The increased value of the QFI in the vicinity of $\Delta_c$ also occurs for eigenstates with energy below $E_5$, as seen for $|\Psi_4\rangle $ in Fig.~\ref{fig:1}(b), and it persists all the way to the ground state, as shown in Fig.~\ref{fig:1}(a).

In addition to the highest value of $F^{(i)}_\Delta$ in Figs.~\ref{fig:1}(b)-(c), shorter peaks are observed. They are correlated with the squeezing of the variance of position, 
\begin{equation}
 \sigma_q^2  = \langle \Psi_i |\hat{q}^2 | \Psi_i \rangle - \langle \Psi_i |\hat{q} | \Psi_i \rangle^2
\end{equation}
shown in Figs.~\ref{fig:1}(d)-(f) for, respectively, the same eigenstates $|\Psi_0\rangle $, $|\Psi_4\rangle $, and $|\Psi_5\rangle $. Thus, peaks in the QFI are mirrored by smaller values of $\sigma_q^2$. Our goal below is to explain, using the analysis of the classical limit of the system in Eq.~(\ref{eq:Ham}), what causes these examples of quantum sensitivity enhancement.

{\em Classical limit and squeezing in position.---} Using Glauber coherent states, $\hat{a}|\alpha \rangle = \alpha |\alpha \rangle$, where $\alpha = q + ip$ and $(q,p)$ are the canonical variables, the classical Hamiltonian associated with the quantum Hamiltonian in Eq.~(\ref{eq:Ham}) reads
\begin{equation}
h_{cl}  \!=\! \langle\alpha|\hat{H}|\alpha\rangle \!=\! -\frac{\Delta}{2} (q^2 + p^2) + \frac{K}{4} (q^2+p^2)^2 - \sqrt{2}P_0 \,q ,
\label{eq:Eqhcl}
\end{equation}
where the classical energy is denoted by $h_{cl}(q,p)=\epsilon$. 

\begin{figure*}[t]
	\centering
\includegraphics[width=0.95\textwidth]{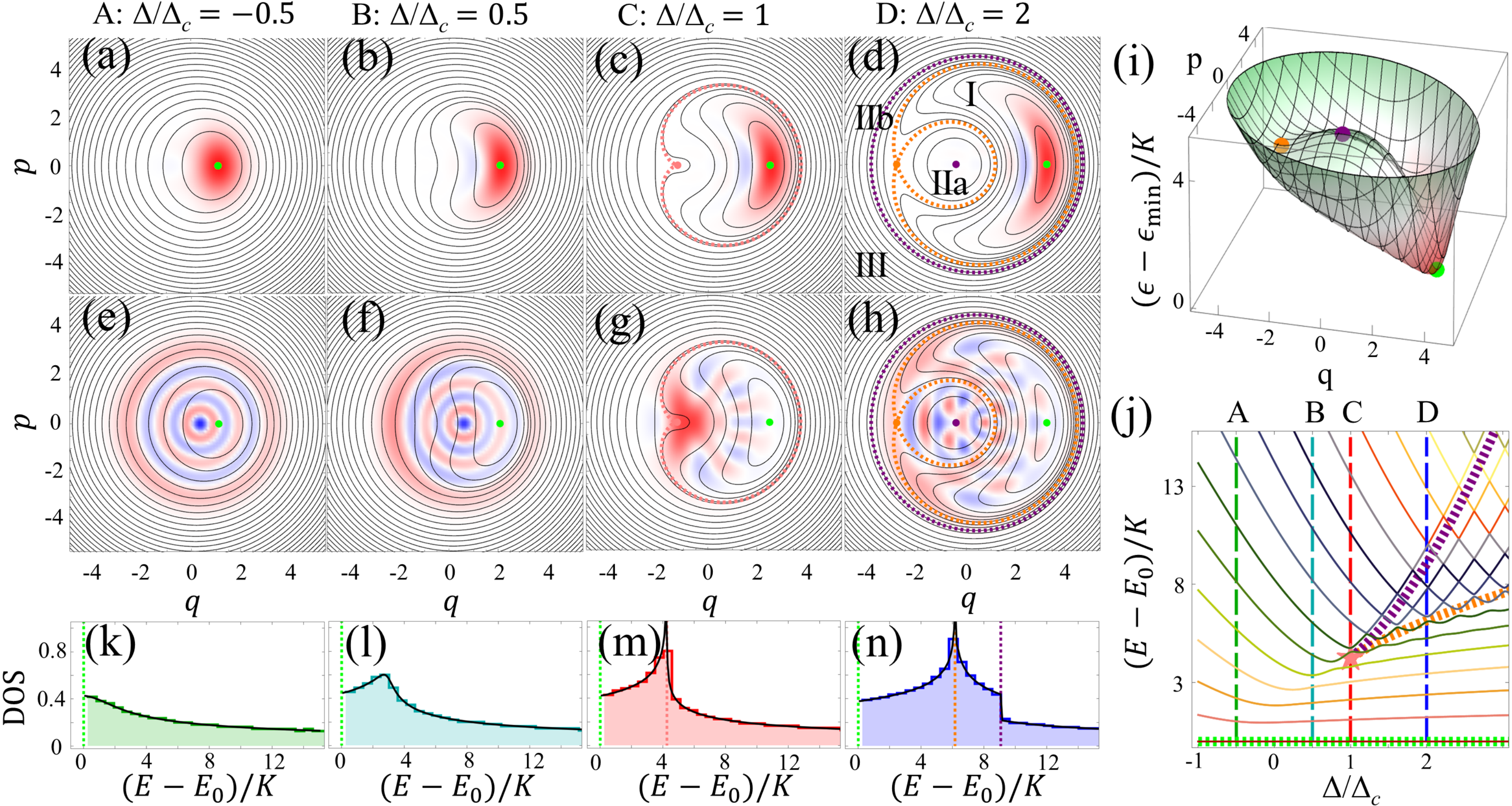}
	\caption{
(a)-(h) Phase space, where the green circle is the global minimum with energy $\epsilon_\text{min}$, the orange circle in (c)-(d), (g)-(h) is the saddle (hyperbolic) point, and the purple point in (d), (h) is the local maximum. The black lines indicate the classical energy shells, the orange dotted line in (c)-(d), (g)-(h) indicates the energy $\epsilon_\text{sad}$ of the saddle point, and the purple dotted line in (d), (h) is the energy $\epsilon_\text{max}$ of the local maximum. 
Panel (i) is a 3D plot of the energy landscape for $\Delta/\Delta_c=2$, as in (d) and (h).
Panel (i): Energy spectrum as a function of $\Delta/\Delta_c$; $P_0/K=3$. Dashed vertical lines indicate the spectrum at A ($\Delta/\Delta_c=-0.5$), B ($\Delta/\Delta_c=0.5$), C ($\Delta/\Delta_c=1$) and D ($\Delta/\Delta_c=2$) used for the analysis in Figs.~\ref{fig:2}(a)-(h). The star at $\Delta/\Delta_c=1$ marks the critical energy $E_{\text{ESQPT}}$ of the first ESQPT. The orange (purple) dashed line is the classical critical energy $\epsilon_\text{sad}$ ($\epsilon_\text{max}$), that coincides with the critical energy $E_{\text{ESQPT}}$ ($E_{\text{step}}$).
Panels (k)-(n): Density of states (shade) and its classical limit (black line); $P_0/K=1000$. The orange (purple) vertical dotted line indicates the critical energy $E_{\text{ESQPT}}$ ($E_{\text{step}}$). 
 }
		\label{fig:2}
	\end{figure*}

The stationary points of this classical system are given by the roots of Hamilton's equations. For any value of the control parameter $\Delta$, the system presents a global minimum, which is indicated with a green dot in the phase-space pictures in Figs.~\ref{fig:2}(a)-(h). 

The black lines in Figs.~\ref{fig:2}(a)-(h) are the classical energy contours. The red and blue shades represent the positive and negative values of the Wigner function of the ground state [Figs.~\ref{fig:2}(a)-(d)] and of the eigenstate $|\Psi_5\rangle$ [Figs.~\ref{fig:2}(e)-(h)].

When the control parameter $\Delta$ coincides with the critical point $\Delta_c=3\sqrt[3]{KP_0^2/2}$, a saddle (hyperbolic) point appears, which is indicated with the orange dot in Figs.~\ref{fig:2}(c,d,g,h). As $\Delta$ increases from zero, the approach of the critical point is accompanied by a deformation of the energy contours, which change from approximately concentric circles in Figs.~\ref{fig:2}(a,e) to shapes that get increasingly elongated in the $p$-direction, as seen in Figs.~\ref{fig:2}(b,f), to finally give space to the onset of the saddle point in Figs.~\ref{fig:2}(c,g). 

These changes in the energy landscape  affect the structures of the eigenstates. The Wigner function of the ground state in Fig.~\ref{fig:2}(c) reveals that the state becomes squeezed in position around the global minimum. In fact, the squeezing in position starts even before the emergence of the saddle point, as noticed in Fig.~\ref{fig:2}(b), and  anticipates its appearance. Similarly, the eigenstate $|\Psi_5 \rangle$ in Fig.~\ref{fig:2}(g) exhibits squeezing around the saddle point. These features explain the minimum value of the variance $\sigma_q^2$ in Figs.~\ref{fig:1}(d)-(f) for $\Delta/\Delta_c \sim 1$.

In addition to the saddle point, the classical system presents a local maximum when $\Delta> \Delta_c$. This point is shown with a purple point in Figs.~\ref{fig:2}(d,h). The orange line associated with the energy of the saddle point, $\epsilon_\text{sad}$, and the purple dotted line at the energy of the local maximum, $\epsilon_\text{max}$, separate four regions in phase space according to their energies [see Fig.~\ref{fig:2}(d)]. They are: the region (I) of lowest energies ($\epsilon < \epsilon_\text{sad}$), the regions (IIa) and (IIb) of intermediate energies ($\epsilon_\text{sad} < \epsilon < \epsilon_\text{max}$), and the outer region (III) of highest energies ($\epsilon>\epsilon_\text{max}$). Figure~\ref{fig:2}(i) provides a 3D illustration of the energy landscape for $\Delta>\Delta_c$ [cf. Figs.~\ref{fig:2}(d,h)].

The squeezing in position for the eigenstates with energy $E \leq \epsilon_{\text{sad}}$ happens because to the emergence of the saddle point is one of the main results of this work. Another crucial finding refers to the peaks in the QFI, whose explanation requires the study of the energy spectrum, as discussed next.

{\em Excited state quantum phase transition.---} In Fig.~\ref{fig:2}(j), we show the eigenvalues $E$ of the Hamiltonian in Eq.~(\ref{eq:Ham}) subtracted from the ground-state energy $E_0$ as a function of the control parameter $\Delta$. We observe that at $\Delta/\Delta_c \sim 1$, the energy levels get closer together in the spectrum region indicated by the orange star. This clustering of the energy levels is reflected in the logarithmic divergence of the density of states shown with numerical data (shade) in Fig.~\ref{fig:2}(m). This divergence is a signature of what became known as ESQPT~\cite{Caprio2008,Cejnar2021} and can be understood with the analysis of the classical limit of the system.

The critical energy of the ESQPT, indicated with the orange star in Fig.~\ref{fig:2}(j), coincides with the energy of the saddle point, $E_\text{ESQPT} \sim \epsilon_\text{sad}$. The dotted orange line that intersects at the saddle point (homoclinic orbit) in Figs.~\ref{fig:2}(c,d,g,h) marks the energy $\epsilon_\text{sad}$ of the separatrix that divides the phase space into two distinct regions. Similarly, $E_\text{ESQPT}$ separates the spectrum in two different regions. The onset of the saddle point causes the logarithmic discontinuity of the level density. Using the Gutzwiller trace formula~\cite{GutzwillerBook}, we derive a semiclassical approximation for the density of states, which is shown with the black line in Fig.~\ref{fig:2}(m).

The clustering of the energy levels at $E_\text{ESQPT}$ is detected by the QFI. The eigenstate $|\Psi_5 \rangle$ has energy very close to $E_\text{ESQPT}$, which explains the peak of its QFI at $\Delta/\Delta_c \sim 1$ in Fig.~\ref{fig:1}(c). Interestingly, the effects of the ESQPT also influence the eigenstates with energies below $E_\text{ESQPT}$, as seen in Figs.~\ref{fig:1}(a)-(b), where the QFI increases near $\Delta/\Delta_c \sim 1$. Additionally, we observe that for these lower-energy states, the highest QFI value shifts toward $\Delta$ values closer to $\Delta_c$ as the system size increases.

Remarkably, the system in Eq.~(\ref{eq:Ham}) exhibits an ESQPT without any sign of a QPT. In several studies of ESQPT, one has seen that the merging of the ground state energy and the first-excited at a QPT propagates to higher energies as the control parameter increases beyond the ground-state critical point~\cite{Caprio2008,Cejnar2021,Khalouf2022QFS}, which is not the case here.

{\em Second excited state quantum phase transition and energy crossings.---} In Fig.~\ref{fig:2}(j), as $\Delta$ increases beyond $\Delta_c$, the critical energy $E_\text{ESQPT}$ grows following the dashed orange line. Figure~\ref{fig:2}(j) also shows a dashed purple line at the energy denoted by $E_\text{step}$, which is associated with the energy of the local maximum, $E_\text{step} \sim \epsilon_\text{max}$, which appears when $\Delta>\Delta_c$. This energy marks the step discontinuity in the density plot displayed in Fig.~\ref{fig:2}(n) and characterizes a second ESQPT.

Between $E_\text{ESQPT}$ and $E_\text{step}$,  the energy levels in Fig.~\ref{fig:2}(j) exhibit avoided crossings that are related to pairs of classical orbits with the same energy, one orbit in region IIa and the other in region IIb of the phase space shown in Fig.~\ref{fig:2}(d). 
The avoided crossings occur when $\Delta/K$ are integers. These values are obtained by identifying the points of degeneracy for the Hamiltonian in Eq.~(\ref{eq:Ham}) when $P_0=0$. In this case, the eigenvalues are $E_n/K=-(\Delta/K +1) n +  n^2$, with $n \in \mathbb{N}$, which leads to degenerate pairs when $(\Delta/K +1)$ are integers~\cite{Iachello2023}. As the external drive is turned on and $P_0 \neq 0$, avoided crossing survive at these same values of the control parameter for the states with energies in $[E_{\text{ESQPT}},E_{\text{step}}]$.

The values of the QFI for the states with energies $E_\text{ESQPT}\lesssim E \lesssim E_\text{step}$ are affected by the energy crossings, resulting in the peaks seen for $\Delta \geq \Delta_c$ in Figs.~\ref{fig:1}(b)-(c). These peaks get higher  as the system size increases.

The energy crossings also influence the squeezing in position. The peaks of the QFI in Figs.~\ref{fig:1}(b)-(c) are correlated with the smaller values of $\sigma_q^2$ in  Figs.~\ref{fig:1}(e)-(f).  This reinforces our claims that quantum sensitivity can be enhanced due to ESQPTs, despite the absence of a QPT, and that the sources of the ESQPTs are  distortions of the phase space, that one can identify by analyzing the classical limit of the system.

{\em Conclusion.---} Our study shows that in experimental platforms with tunable control parameters, such as exciton-polariton condensates and superconducting circuits,  excited state quantum phase transitions can be externally induced and exploited for quantum sensing. The clustering of eigenvalues, energy crossings, and changes in the structures of the eigenstates associated with these transitions amplify the quantum Fisher information and simultaneously squeeze the uncertainty in position, resulting in enhanced quantum sensitivity.  Our approach paves the way for the development of remotely tunable platforms for sensing and metrology.

{\em Acknowledgements.---} The authors acknowledge support from the National Science Foundation Engines Development Award: Advancing Quantum Technologies (CT) under Award Number 2302908. JCC, VSB and LFS also acknowledge partial support from the National Science Foundation Center for Quantum Dynamics on Modular Quantum Devices (CQD-MQD) under Award Number 2124511. MABM acknowledges financial support from CONAHCYT CBF2023-2024-1765, DCBI UAM-I PEAPDI 2024, and DAI UAM PIPAIR 2024 projects. FPB received funding from the Grant PID2022-136228NB-C21 funded by MICIU/AEI/
10.13039/501100011033 and, as appropriate, by "ERDF A way of making Europe", by
"ERDF/EU," by the "European Union," or by the "European Union
NextGenerationEU/PRTR." Computing resources supporting this work were partly provided by the CEAFMC and Universidad de Huelva High Performance Computer (HPC@UHU) located in the Campus Universitario "El Carmen" and funded by FEDER/MINECO project UNHU-15CE-2848.



\bibliography{main}

\end{document}